\documentclass{ifacconf}

\usepackage{natbib}        
\usepackage{amsmath,amsfonts}
\usepackage{graphicx,xcolor}
\usepackage{enumitem}
\usepackage{subcaption}
\usepackage[hyphens]{url}
\newtheorem{definition}{Definition}{}
{}
\newtheorem{remark}{Remark}

\newcommand{\R}{\mathbb{R}}
\newcommand{\N}{\mathcal{N}}
\newcommand{\Nb}{\mathfrak{N}}

\newcommand{\G}{\mathcal{G}}
\newcommand{\norm}[1]{\left\|#1 \right\|}
\begin{document}
\begin{frontmatter}

\title{History Data Driven Distributed Consensus in Networks} 

\thanks[footnoteinfo]{V. Renganathan \& A. Fontan contributed equally. This project has received funding from the European Research Council (ERC) under the European Union’s Horizon 2020 research and innovation program under grant agreement No 834142 (Scalable Control).}

\author[First]{Venkatraman Renganathan} 
\author[Second]{Angela Fontan} 
\author[Third]{Karthik Ganapathy}

\address[First]{Department of Automatic Control - LTH, Lund University, Lund, Sweden (e-mail: venkatraman.renganathan@control.lth.se).}
\address[Second]{Division of Decision and Control Systems, KTH Royal Institute of Technology, Stockholm, Sweden (e-mail: angfon@kth.se)}
\address[Third]{Department of Mechanical Engineering, The University of Texas at Dallas, Richardson, TX, USA (e-mail: karthik.ganapathy@utdallas.edu)}

\begin{abstract}                
The association of weights in a distributed consensus protocol quantify the trust that an agent has on its neighbors in a network. An important problem in such networked systems is the uncertainty in the estimation of trust between neighboring agents, coupled with the losses arising from mistakenly associating wrong amounts of trust with different neighboring agents. We introduce a probabilistic approach which uses the historical data collected in the network, to determine the level of trust between each agent.
%
Specifically, using the finite history of the shared data between neighbors, we obtain a configuration which represents the confidence estimate of every neighboring agent’s trustworthiness.
Finally, we propose a History-Data-Driven (HDD) distributed consensus protocol which translates the computed configuration data into weights to be used in the consensus update. The approach using the historical data in the context of a distributed consensus setting marks the novel contribution of our paper.
\end{abstract}


\end{frontmatter}

\section{Introduction}
We study the problem of consensus in a multi-agent system in the presence of untrustworthy agents in this paper.
Many cooperative tasks involving networked agents require them to utilize distributed consensus protocols to coordinate agreement on certain quantities of interest, with applications such as formation control in robotics (\cite{Fax2004Cooperative,Ren2008Distributed}), agreement seeking in opinion dynamics (\cite{Hegselmann2002Opinion,Blondel2009Krause,Fontan2021Role}), or cyber-networks comprising of many interconnected smart entities which relies on distributed consensus protocols for efficient operations (\cite{venki_tcst}). 
However, \cite{Pasqualetti2012Consensus} showed that the distributed nature of networks opens up many attack points for malicious attackers rendering them vulnerable. 
This work considers the situation where well-behaving agents (called ``cooperative'' in our notation) in a network seek to achieve consensus in the presence of ``untrustworthy'' agents (called ``potentially non-cooperative'' in our notation) using the inference from the past interactions with their neighbors.

A related problem is that of consensus in unreliable networks, which has been largely studied in the literature, see e.g., \cite{leslie_byzantine, peleg_robots, Sundaram2008Distributed,leblanc_wmsr,Saldana2017Resilient,Tempo2018Resilient}, and specifically resilient consensus protocols, such as the W-MSR protocol by \cite{leblanc_wmsr}, have been developed in the recent past to guarantee resiliency by intelligently constructing a nonlinear consensus update. In general, the update rule of these distributed consensus protocols depends upon the current time step information obtained from all the neighboring agents in the network. An exception is the protocol for resilient consensus proposed in \cite{Saldana2017Resilient}, named SW-MSR, which extends the classical W-MSR algorithm by introducing a sliding window approach that allows the agents to store the values received from their neighbors at the previous $T$ time steps.
While the resilient consensus literature imposes an assumption on the connectivity and the total number of non-cooperative (also called non-reliable or malicious) agents in the network, the recent work by \cite{Yemini2021Trust} departs from such assumptions and uses the notion of trust in order to maintain consensus in a networked system in the presence of malicious agents.

Our aim is to design a distributed consensus protocol that enables each agent to estimate the trustworthiness of its neighbors, represented by a (normalized) non-negative value in $[0,1]$, where $0$ (resp., $1$) indicates that the corresponding agents do not trust (resp., trust fully) each other, 
with the idea that an agreement should be reached only between agents whose trust in each other is nonzero.
Similar to \cite{Yemini2021Trust}, in this work we do not impose any structural or connectivity assumption on the network or any assumption on the total number of potentially non-cooperative agents, but we consider the observed history at the previous $T$ time steps to estimate trust between each agent in the network.
Such an approach offers a paradigm shift from considering memory-less update in distributed consensus protocols, which do not offer the debugging capabilities of getting to know \textit{when and where} an intentional attack or a fault happened in the network, thereby possibly lacking the retrospecting ability to analyze for anomalies. On the other hand, it is not practical for every agent in a network to have infinite book-keeping abilities to store the shared values of its neighbors information to analyze for any anomalies. However, given a finite memory resource is made available for agents in a network, distributed consensus algorithms can be reinforced with retrospecting abilities to enhance the quality of the decisions that they make and mimic the trust-based decision-making behavior of humans in a distributed setting.

Under the local information model setting, the protocol we propose is related to the bounded confidence models in opinion dynamics (see for instance \cite{Hegselmann2002Opinion}) where each agent updates its state (analogous to its opinion) based only on the states of agents that are within a certain confidence range of its own, enforcing the idea that only trustworthy agents (here intended as agents with similar opinions) can influence each other. Moreover, inspired by \cite{Lorenz2009Heterogeneous,Liang2013Opinion,Morarescu2011Decaying}, we assume that the confidence bounds are heterogeneous (i.e., agent-dependent) and time-dependent. The problem of consensus in networks with random weighting matrices was studied in \cite{ali_tac, ali_allerton}. We refer to a closely aligned idea that appeared in \cite{yu_mint, yu_dro}, where malicious nodes were identified in an uncertain network with high confidence and removed. We consider extending a similar idea as \cite{yu_dro} for designing a distributed consensus protocol using an history data-driven approach. Specifically, at each time step, we use the available finite historical data to estimate the first two moments of an unknown distribution governing the true nature (called ``configuration'' in our notation) of an agent's neighbors. 

\emph{Statement Of Contributions:} We propose a novel history data-driven distributed consensus protocol for networks. Specifically, our main contributions are as follows:
\begin{enumerate}
    \item We model the true nature of neighboring agents of an agent in a network as a random vector (which we term as the ``configuration'' of neighbors), and we learn the parameters governing its true but unknown distribution from the collected historical data.
    
    \item We translate the trustworthiness that resulted from the neighbor configuration into weights and propose a new History-Data-Driven (HDD) distributed consensus protocol for networks.
    
    \item We demonstrate by means of numerical simulation that our proposed design effectively models the neighbor configuration from the historical data, and arrives at a trust-based consensus\footnote{Notion introduced in Definition~\ref{def:trusted_consensus}.}. 
\end{enumerate}
The rest of the paper is organized as follows: The preliminaries of the consensus protocol and the definition of a neighbor configuration are established in section \ref{sec_prob_formulation}. In section \ref{sec_estimate_p_hat}, the empirical estimation of the configuration parameters from the past historical data is discussed. The proposed HDD distributed consensus protocol is presented in section \ref{sec_hdd_protocol} along with the effect of parameter variations. Our proposed algorithm is then demonstrated in section \ref{sec_numerical_example}. Finally, the paper is closed in section \ref{sec_conclusion} with a summary and research directions for the future.

\section*{Notation \& Preliminaries}
We denote the set of real numbers, integers, non-negative real numbers and non-negative integers by $\mathbb{R}, \mathbb{Z}, \mathbb{R}_{\geq 0}, \mathbb{Z}_{\geq 0}$ respectively. The operator $\backslash$ denotes the set subtraction. The cardinality of the set $M$ is denoted by $|M|$ and its $i$\textsuperscript{th} element by $\{M\}_{i}$. The $i$\textsuperscript{th} element of a vector $x$ is denoted by $[x]_{i}$ or simply $x_{i}$ and the Euclidean norm of $x$ is denoted by $\left \Vert x \right \|_{2}$ or simply $\left \Vert x \right \|$. A vector in $\mathbb{R}^{n}$ with all its elements being ones is denoted by $\mathbf{1}_{n}$. The $j$\textsuperscript{th} column of a matrix $A$ is denoted by $A_{j}$. An element in $i$\textsuperscript{th} row and $j$\textsuperscript{th} column of matrix $A$ is denoted by $A_{ij}$. The uniform distribution between $a, b \in \mathbb{R}$, $a < b$ is denoted by $U[a,b]$.
\section{Problem Formulation} \label{sec_prob_formulation}
\subsection{Consensus Dynamics of a Network} \label{subsection_update_model}
Consider a network having $N$ agents whose connectivity is modeled via an undirected and connected graph $\mathcal{G}=(\mathcal{V},\mathcal{E})$, where $\mathcal{V}$ represents the set of agents with $|\mathcal{V}| = N$.  A set of time-invariant communication links amongst the agents is represented using $\mathcal{E} \subset \mathcal{V} \times \mathcal{V}$. We associate with each agent $i \in \mathcal{V}$ a state $x_{i}(t) \in \mathbb{R}$ at time $t \in \mathbb{Z}_{\geq 0}$. 
Let the set of inclusive neighbors be defined as $\mathcal{J}_{i} = \N_{i} \cup \{i\}$, where $\N_{i} = \{j \in \mathcal{V} : (j,i) \in \mathcal{E}\}$ is the neighbor set of agent $i$, whose states are available to agent $i$ via communication links. The degree of $i$ is denoted as $d_{i} = |\N_{i}|$, and every agent is assumed to have access to its own state at any time $t$. At any time $t$, each agent updates its own state based on its current state and the states of its neighboring agents according to a prescribed memory-less update rule 
\begin{equation} \label{eqn_usual_consensus_update}
x_{i}(t+1) = f_{i}(x_{j}(t)), \, \, j \in \mathcal{J}_{i}, \ i \in \mathcal{V}.
\end{equation}
Typical distributed consensus protocols of the form \eqref{eqn_usual_consensus_update} involve associating a weight corresponding to all inclusive neighbors $j \in \mathcal{J}_{i}$ and using it in the consensus update. In this work, we consider the weighted averaging type of update protocols, namely $x(t+1) = W(t) x(t)$ with $x(t) = [x_1(t) \dots x_N(t)]^{\top}$, that is, 
\begin{equation} 
    x_{i}(t+1) = \sum_{j \in \mathcal{J}_{i}} w_{ij}(t) x_{j}(t), \quad i \in \mathcal{V}, \label{eqn_dcp}
\end{equation}
where $W(t)$ is an element-wise non-negative time-varying weighting matrix with its entries $w_{ij}(t)\ge 0$ modeling the trustworthiness associated by agent $i$ on its inclusive neighboring agents $j \in \mathcal{J}_{i}$ at each time $t$. 
\cite{olshevsky2009convergence} showed that, under assumptions on the graph (such as connectivity of $\G$) and on the weights of $W(t)$ (e.g., weights chosen according to a convex combination, making $W(t)$ a stochastic matrix)%
, an asymptotic consensus value is guaranteed by \eqref{eqn_dcp}, that is, $\exists\, c \in \R$ s.t. $\lim_{t\to\infty} x(t)=c \mathbf{1}_N$. 

In this work we consider the situation in which some agents may not follow the update rule \eqref{eqn_dcp}. To this end, we introduce the notion of cooperative and (potentially) non-cooperative agents.
\begin{definition}\label{def:normal-faulty}
An agent $i \in \mathcal{V}$ is said to be \textit{cooperative} if it updates its state based on \eqref{eqn_dcp}. It is said to be \textit{(potentially) non-cooperative}%
\footnote{For instance, an agent $i \in \mathcal{V}$ can act \emph{non-cooperatively} by applying random update function $f^{\prime}_i (\cdot)$ other than \eqref{eqn_dcp} at all time-steps.}, otherwise. 
\end{definition}
\begin{definition}\label{def:trusted_consensus}
An agent $i \in \mathcal{V}$ is said to be in a \textit{trust-based consensus} with a set of identified trusted neighbors $j \in \overline{\N}_{i} \subseteq \mathcal{N}_{i}$ if $\displaystyle\lim_{t\to \infty} \norm{x_i(t) - x_j(t)} =0$ for all $j \in \overline{\N}_{i}$.
\end{definition}
The intuition is that if the cooperative agents manage to correctly identify and distrust non-cooperative agents, effectively trusting only (a subset of) their neighbors, 
then the sequence $\{x_i(t)\}_{t\ge 0}$ is convergent for all cooperative agents $i$, i.e., $x_i(t)\to x_i^\ast\in \R$, and either the agents reach an agreement, i.e., $x_i^\ast=x_j^\ast$ for all $i,j\in \mathcal{V}$, or clustering, i.e., $x_i^\ast=x_j^\ast$ for all $i,j$ belonging to the same cluster. However, a smart non-cooperative  agent, if undetected, may act as a leader and be followed by a set of cooperative agents (whose corresponding sequence $\{x_i(t)\}_{t\ge 0}$ then, in that case, need not be convergent).

\subsection{Availability of Historical Data}
When a memory-less distributed protocol like \eqref{eqn_dcp} is used by the cooperative agents $i \in \mathcal{V}$ under the setting where some agents \emph{might} be non-cooperative, the resulting asymptotic consensus can be easily manipulated by some smart adversaries. Under this setting, the mechanism we propose for each agent with uncertain information on the nature of its neighbors, is to observe the neighbors' shared data for a certain period of time, arrive at an estimate of the neighbors' trustworthiness, and subsequently use it in its update to arrive at a consensus. To facilitate a tractable problem formulation and to make the resulting consensus algorithm suitable for dynamic implementation, we consider a finite historical data of length $T \in \mathbb{Z}_{\ge 0}$ updated in a rolling horizon fashion\footnote{Future work will seek to understand the behaviour with growing history over time.}. At all time steps $t$, every agent $i \in \mathcal{V}$ is \emph{assumed} to have access to the history of its own values and to its neighboring agents' values $x_{j}(t), j \in \mathcal{N}_{i}$ for the past $T$ time-steps. 
That is, with $\kappa_{t, T} = \{t-l\}^{T-1}_{l=0}$, we have
\begin{gather*}
\mathfrak{X}^{T}_{i,j}(t) = \left\{ x_{j}(k) \, \mid \, j \in \mathcal{N}_{i}, \, k \in \kappa_{t, T} \right\}, \\
\mathfrak{X}^{T}_{i,i}(t) = \left\{ x_{i}(k) \mid k \in \kappa_{t, T} \right\},\\
\mathfrak{X}^{T}_{i,\mathcal{N}_{i}} (t)= \left\{ \mathfrak{X}^{T}_{i,j}(t) \, \mid \, j \in \mathcal{N}_{i} \right\}. 
\end{gather*}
Thus, the main purpose of this work is
\textit{(i) to design a protocol that allows each cooperative agent $i\in \mathcal{V}$ to estimate the trustworthiness of its neighbors at each time step $t$, given the history $\mathfrak{X}^{T}_{i,\mathcal{N}_{i}}(t)$ and $\mathfrak{X}^{T}_{i,i}(t)$ for the past $T$ time steps; (ii) to study the role of the estimated trustworthiness in solving the trust-based consensus problem for the cooperative agents, despite the presence of non-cooperative agents in the network.}



\section{Set Membership Based Empirical Estimation of Configuration}\label{sec_estimate_p_hat} 
In this section, we describe how to learn the trustworthiness of neighbors, given their $T$ time steps historical data. For each agent $i\in \mathcal{V}$, the \textit{configuration} of neighbors $\mathcal{N}_{i}$ at time $t$, denoted by $\pi^{i}_{t} \in [0,1]^{d_{i}}$, encodes the degree of trustworthiness of every neighbor $j \in \mathcal{N}_{i}$. A neighbor $j \in \mathcal{N}_{i}$ is said to be completely trustworthy or not trustworthy if $\left[\pi^{i}_{t}\right]_{j}$ equals to 1 or 0, respectively, and any value in $[0,1]$ defines its degree of trustworthiness. 
Similarly, we define $\Bar{\pi}^{i}_{t} = \mathbf{1}_{d_{i}} - \pi^{i}_{t}$ to be the configuration representing the degree of non-cooperativeness of the neighbors at time $t$. Note that $\pi^{i}_{t}$ is a random vector where the $j\textsuperscript{th}$ entry of $\pi^{i}_{t}$ corresponding to the neighbor $j \in \mathcal{N}_{i}$ is supported on a compact interval $[0,1]$. Further, $\pi^{i}_{t} \sim \mathbb{P}^{i}_{t}$ with $\mathbb{P}^{i}_{t}$ denoting the true but \emph{unknown} distribution of the $\pi^{i}_{t}$ supported on a compact set $[0,1]^{d_{i}}$. Let $\mu^{i}_{t} \in \mathbb{R}^{d_{i}}$ and $\Sigma^{i}_{t} \in \mathbb{R}^{d_{i} \times d_{i}}$ denote the true mean and covariance respectively associated with $\mathbb{P}^{i}_{t}$. Though in reality $\mathbb{P}^{i}_{t}$ is not readily available, it can be estimated from data, that is, using the $T$ time steps history data, it is possible to form an empirical distribution $\hat{\mathbb{P}}^{i}_{t}$. Let us denote the mean and the covariance of $\hat{\mathbb{P}}^{i}_{t}$ by $\hat{\mu}^{i}_{t}$ and $\hat{\Sigma}^{i}_{t}$, respectively. Here, $\left[\hat{\mu}^{i}_{t}\right]_{j}$ is agent $i$'s estimated trustworthiness at time $t$ about the neighboring agent $j \in \mathcal{N}_{i}$ given its past $T$ time steps historical data. We propose a set membership based approach to estimate the parameters of the empirical configuration distribution $\hat{\mathbb{P}}^{i}_{t}$ given the historical data $\mathfrak{X}^{T}_{i,i}(t)$ and $\mathfrak{X}^{T}_{i,\mathcal{N}_{i}}(t)$. We base the following discussion on the presumption that a neighbor $j \in \mathcal{N}_{i}$ is believed to be more trustworthy by agent $i$, if $j$'s values are in the desired vicinity of agent $i$'s value throughout the considered past. 

\subsection{The $\epsilon$-Neighborhood Based Set Membership}
To define a set membership based estimation, we require a set of confidence neighborhoods for all the past $T$ time steps. Thus, for all $k \in \kappa_{t, T}$, the confidence neighborhood around the $x_{i}(k)$ is defined as, 
\begin{equation}
    \mathcal{B}_{x_{i}(k)}(\epsilon_{i, k}) = \left\{ y \in \mathbb{R} \mid \left \Vert y - x_{i}(k) \right \|_{2} \leq \epsilon_{i, k} \right\},
\end{equation}   
where $\epsilon_{i, k} > 0$ is the confidence bound for agent $i$ at time $k$. To value the recent past more than the distant past
, we assume that at each time step $t$ agent $i \in \mathcal{V}$ is free to choose a decreasing sequence of confidence bounds $\epsilon_{i,k},\,\forall k \in \kappa_{t,T}$ as follows:
\begin{equation} \label{eqn_eps_seq}
    \epsilon_{i, t-(T-1)} > \dots > \epsilon_{i, t-2} > \epsilon_{i, t-1} > \epsilon_{i, t} > 0.
\end{equation}
Using the confidence neighborhoods $\mathcal{B}_{x_{i}(k)}(\epsilon_{i, k})$, and the information sets $\mathfrak{X}^{T}_{i,i}(t), \mathfrak{X}^{T}_{i,\mathcal{N}_{i}}(t)$, we define the set membership counter for all time steps $k \in \kappa_{t, T}$ as follows,
\begin{equation}
    \mathfrak{N}^{i}_{k} = \left\{ j \in \mathcal{N}_{i} \mid x_{j}(k) \in \mathcal{B}_{x_{i}(k)}(\epsilon_{i, k}) \right\}.
\end{equation}
Here, $\mathfrak{N}^{i}_{k} \subseteq \N_{i}$ accounts for the neighbors $j \in \mathcal{N}_{i}$ who share their values in the vicinity of the agent $i$, at time step $k$ of the past history.
It is possible that at time $k\in \kappa_{t, T}$, the set $\mathfrak{N}^{i}_{k}$ may turn out to be empty, or equal to $\N_{i}$.

\subsection{Estimating Parameters of Configuration Distribution}
Now, we illustrate how each agent $i\in \mathcal{V}$ estimates the trustworthiness of its neighbors at each time instant $t$.
First, a frequency counter and a discounted importance vector are defined for each agent as follows.
A frequency counter $\mathcal{C}^{i}_{j}(t)$ records, at each time step $t$, the time indices $k\in \kappa_{t, T}$ where the neighbor $j \in \mathcal{N}_{i}$ belonged to $\mathfrak{N}^{i}_{k}$:
\begin{equation*}
    \mathcal{C}^{i}_{j}(t) = \left\{ k \in \kappa_{t, T} \mid j \in \mathfrak{N}^{i}_{k} \right\}\quad j \in \mathcal{N}_{i}.
\end{equation*}
A discounted importance vector $\mathfrak{d}^{i}_{j}\in \mathbb{R}^{T}$ qualitatively captures how a neighboring agent $j \in \mathcal{N}_{i}$ behaved with respect to the agent $i$, by valuing the recent past more than the distant past using a discount factor $\nu_{i, t} \in (0,1)$:
\begin{equation*}
    \left[\mathfrak{d}^{i}_{j}\right]_{k} = \begin{cases} \nu^{t-k}_{i, t}, & \text{if } k \in \mathcal{C}^{i}_{j}(t), \\
    0, & \text{if } k \notin \mathcal{C}^{i}_{j}(t).
    \end{cases}
\end{equation*}
Finally, the estimated mean $\hat{\mu}^{i}_{t}$ and estimated covariance $\hat{\Sigma}^{i}_{t}$ at time $t$ are computed as:
\begin{gather}
\left[\hat{\mu}^{i}_t\right]_{j} =\frac{1}{T} \sum_{k \in \kappa_{t, T}} \left[\mathfrak{d}^{i}_{j}(t)\right]_{k}, \quad j \in \mathcal{N}_{i}, \label{eqn_estim_mean}
\\
\hat{\Sigma}^{i}_t = \frac{1}{T-1} \sum_{k \in \kappa_{t, T}} \left(\mathcal{D}^{i}_{k} - \hat{\mu}^{i}_t\right) \left(\mathcal{D}^{i}_{k} - \hat{\mu}^{i}(t)\right)^{\top}, 
\label{eqn_estim_covar}
\end{gather}
where the variability matrix\footnote{Several choices for defining the matrix $\mathcal{D}^{i}$ exist other than \eqref{eqn_variable_matrix}.} $\mathcal{D}^{i} \in \mathbb{R}^{d_{i} \times T}$ is given by
\begin{equation} \label{eqn_variable_matrix}
    \mathcal{D}^{i}_{jk} = 
    \frac{\norm{x_{i}(k) - x_{j}(k)}}{1+\norm{x_{i}(k) - x_{j}(k)}}, \quad j \in \mathcal{N}_{i}, k \in \kappa_{t, T}.
\end{equation}
Then an estimate\footnote{This style of inferring the trustworthiness has the potential of being vulnerable with smarter adversaries as they can manipulate the estimated parameters $\hat{\mu}^{i}_{t}$ and $\hat{\Sigma}^{i}_{t}$ to render them off from their respective true values $\mu^{i}_{t}$ and $\Sigma^{i}_{t}$. Future research will seek to address this using the distributionally robust stochastic program (DRSP) model described in \cite{delage2010distributionally}.} of trustworthy configuration is $\hat{\pi}^{i}_{t} = \hat{\mu}^{i}_{t}$, and subsequently $\hat{\bar{\pi}}^{i}_{t} = \mathbf{1}_{d_{i}} - \hat{\pi}^{i}_{t}$ would be an estimate of non-cooperative configuration. In the next section, we elucidate how an agent $i \in \mathcal{V}$ can use the inferred trustworthiness of its neighbors $j \in \mathcal{N}_{i}$ to update its value.


\section{Design of An Historical Data-Driven Distributed Consensus Protocol} \label{sec_hdd_protocol}
In this section, we use the obtained trustworthiness information of the neighbors of an agent to design a distributed consensus protocol. Following the definition of HDD protocol, we discuss the effect of various parameter variations on the consensus obtained using HDD protocol.

\subsection{An HDD Distributed Consensus Protocol}
For every collaborative agent $i \in \mathcal{V}$, $\hat{\mu}^{i}_{t}$ denotes the estimated trustworthiness of its neighbors given their past historical data. Further, every collaborative agent $i$ has to trust itself completely at all time steps. Therefore, at each time step $t$, we form the augmented trust vector $z^{\dagger}_{i}(t) \in \mathbb{R}^{\left | \mathcal{J}_{i} \right \vert}$ as
\begin{equation} \label{eqn_aug_trust}
    z^{\dagger}_{i}(t) = \begin{bmatrix}
    \hat{\mu}^{i}_{t} \\ 1
    \end{bmatrix}, \quad \text{since } \mathcal{J}_{i} = \mathcal{N}_{i} \cup \{i\}.
\end{equation}
Then, every collaborative agent $i \in \mathcal{V}$ updates its states using the following proposed History-Data-Driven (HDD) distributed consensus protocol as follows
\begin{equation} \label{eqn_hdd_update}
    x_{i}(t+1) = \sum_{j \in \mathcal{J}_{i}} \underbrace{\frac{[z^{\dagger}_{i}(t)]_{j}}{\left\Vert z^{\dagger}_{i}(t) \right \|_{1}}}_{:=w_{ij}(t)} x_{j}(t),
\end{equation}
where the weights $w_{ij}(t)\in [0,1],  \forall i \in \mathcal{V}, j \in \mathcal{J}_{i}$, and $\sum_{j\in \mathcal{J}_i} w_{ij}(t) = 1$. Moreover, $w_{ij}(t) = 0$ if and only if $j \notin \mathcal{J}_{i}$ or $j \notin \Nb^i_k$ for all $k\in \kappa_{t,T}$.
\begin{remark}
The weights $w_{ij}(t)$ given by \eqref{eqn_hdd_update} translates the trustworthiness information of neighboring agents into weights for the distributed consensus update rule. The HDD protocol is actually a nonlinear consensus update, such as the W-MSR protocol, as the weight $w_{ij}(t)$ computed by agent $i$ for its neighbor $j \in \mathcal{N}_{i}$ with an informed choice of the parameters $T, \epsilon_{i,k}$ and $\nu_{i,t}$ may turn out to be zero based on the inference using the historical data, meaning that at time $t$, agent $i$ neglects neighbor $j$'s contribution. 
\end{remark}
\begin{remark}
Following the proof of Proposition~1 present in \cite{Morarescu2011Decaying}, it is possible to prove that, no matter the choice of $T\ge 1$ and $\nu_{i,t}$, the sequence $\{x_i(t)\}_{t\ge 0}$ is convergent for all cooperative agents $i$, that is, $x_i(t)\to x_i^\ast$, if, for each cooperative agent $i$, the confidence bounds $\{\epsilon_{i,k}\}_{k\in \kappa_{t,T}}$ at each instant $t$ are chosen so that $\epsilon_{i,t-T+1}$ is time-decaying and eq.~\eqref{eqn_eps_seq} is satisfied\footnote{Future research will explore this direction and seek to address provable convergence to \textit{finite-time} trust-based consensus.}.
Possible choices are, for instance, $\epsilon_{i,t-T+1}= R_i e^{-\rho_i (t-T+1)}$ or $\epsilon_{i,t-T+1}= R_i \rho_i^{t-T+1}$, where $R_i\in \mathbb{R}_{\ge 0}$ and $\rho_i \in (0,1)$.
\end{remark}

\subsection{Effects of Parameter Variations} 
The design parameters of our algorithm are $T\in \mathbb{Z}_{\geq 0}$ ($T$ finite), $\{\nu_{i,t}\}_{i \in \mathcal{V},t \in \R_+}$, and $\{\epsilon_{i,k}\}_{i \in \mathcal{V},k \in \kappa_{t,T}}$. The parameter $\nu_{i, t}\in (0, 1)$ can be regarded as the forgetting factor for an agent $i \in \mathcal{V}$ at time $t$ and thus influences how much an agent is willing to remember its neighbors' past interactions from the time $t$ given the history length $T$ (defined respecting the available memory constraints). For instance, $\nu_{i, t}$ closer to $1$ indicates that the agent emphasizes its recent past interactions with its neighbors more and hence its forgetfulness decreases rather slowly over time. On the other hand, $\nu_{i, t}$ closer to $0$ indicates that the agent forgets quickly. The next parameter confidence bound $\epsilon_{i, k}$ at time $k$ is the agent $i$'s freedom to choose its desired vicinity area around its values to value its neighbors appropriately. For instance, a cautious agent $i \in \mathcal{V}$ would tend to have a small $\epsilon_{i, k}$ even at its distant past, while a relaxed agent may tend to choose a generous $\epsilon_{i, k}, \forall k \in \kappa_{t,T}$. Given that $\mathfrak{N}^{i}_{k}$ for an agent $i \in \mathcal{V}$ directly depends upon the $\{\epsilon_{i, k}\}_{k \in \kappa_{t,T}}$ and which in turn defines the $\hat{\mu}^{i}_{t}$, it is clear that the resulting consensus is directly impacted by the choice of the confidence bound that an agent chooses according to its behavioural aspects. Though the efficacy of the proposed update protocol is limited upon the memory constraint defining the parameter $T$, its freedom in the design of other design parameters makes it both an interesting and powerful consensus protocol. 

\begin{remark}
The collected historical data can be used to predict the neighbors value using machine learning techniques and if the neighbor shares a value closer to the predicted value, agent $i$ can allocate an higher trust to value their contribution more and thereby define \emph{data-driven predictive consensus} protocol. Another variation to HDD protocol would be to remove neighbors whose trustworthiness fall below a specified trust-threshold and subsequently using only the remaining neighbors values. Future work will seek address above variations and to investigate adaptive designs of discount factor $\nu_{i, t}$ and  confidence bounds $\epsilon_{i, k}, \forall k \in \kappa_{t,T}$.
\end{remark}
\section{A Numerical Example} \label{sec_numerical_example}
In this section, we elaborate the simulation results that we performed to demonstrate our proposed HDD distributed consensus protocol. We considered an undirected graph $\mathcal{G}=(\mathcal{V}, \mathcal{E})$ with $|\mathcal{V}| = N = 13$ agents. Further, the first 10 agents were assumed to be cooperative, and are denoted by $\mathcal{V}_{c} = \{1, \dots,10\}$, and the rest $\mathcal{V}_{nc} = \mathcal{V} \backslash \mathcal{V}_{c}$ to be non-cooperative. The cooperative nodes $i \in \mathcal{V}_{c}$ were randomly connected with probability $p=0.4$ (here, $p$ denotes the probability of an edge between two nodes), while each non-cooperative node $i \in \mathcal{V}_{nc}$ was connected to all cooperative nodes.
All agents were given a random fixed history of data generated for a considered history of length $T$. The HDD protocol was demonstrated for a total of $T_t = 200$ time steps with the new state trajectories of neighboring agents being used to update the historical data in a rolling horizon fashion. Each agent $i \in \mathcal{V}$ was given random confidence bounds (decreasingly sorted) for all $t$ where each $\epsilon_{i,k}$ was drawn from $U[\underline{\epsilon},\overline{\epsilon}]$ with the lower limit set to a constant value, $\underline{\epsilon} = 0.01$ and the upper limit was varied as $\overline{\epsilon} \in \{0.5,1,1.5\}$ to observe different behaviours. Every agent $i \in \mathcal{V}$ was given the same discount factor $\nu_{i,t}=\nu \in (0,1)$ at a given time step $t$. 
The HDD protocol was executed by varying one of the parameters $T$, $\nu$, $\overline{\epsilon}$ 
while keeping the rest fixed.

\begin{figure*}
\begin{subfigure}{.5\textwidth}\centering
  \includegraphics[width=\textwidth]{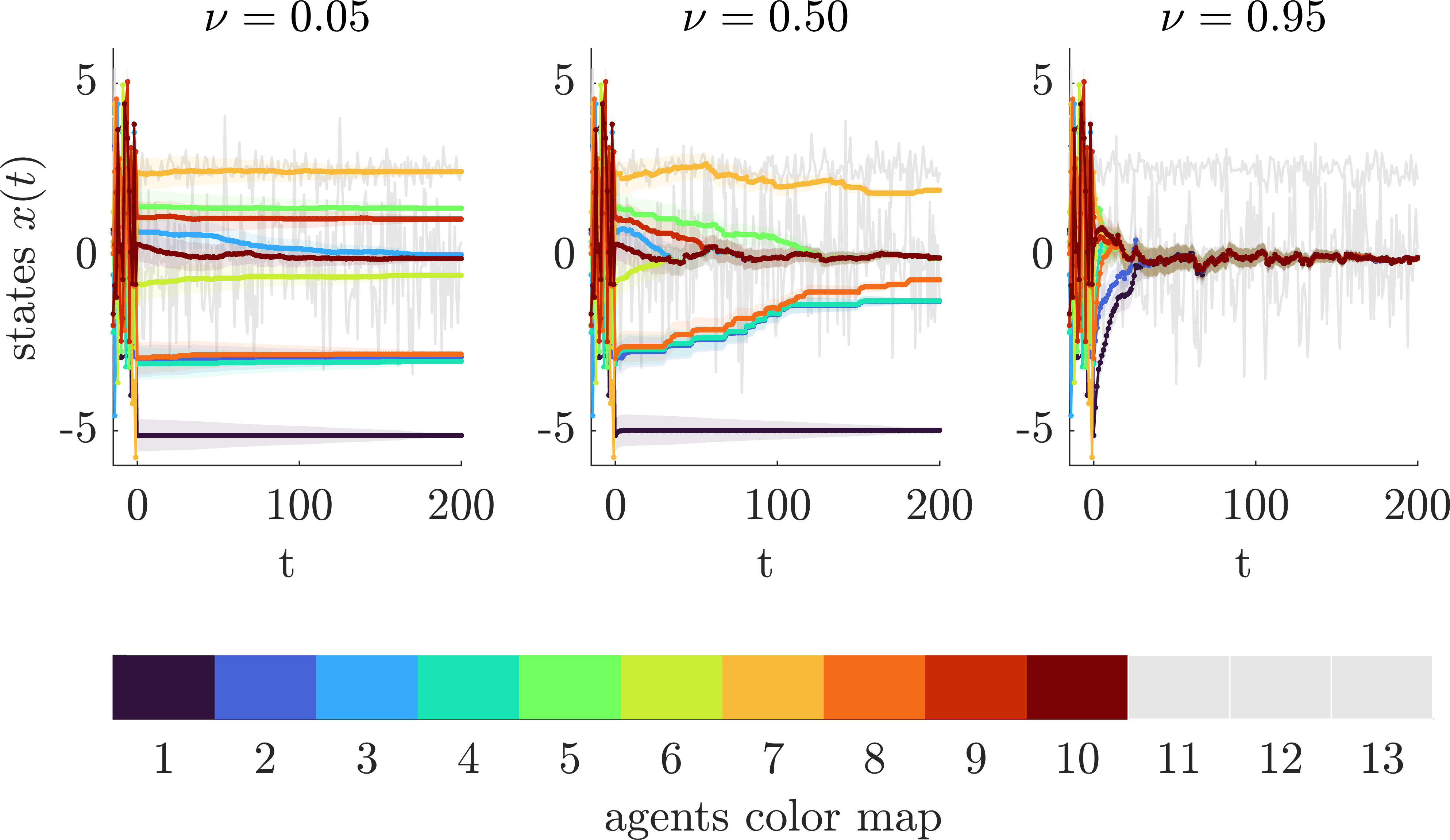}
  \caption{HDD protocol~\eqref{eqn_hdd_update} with $T = 15$, $\{\epsilon_{i,k}\}_{k \in \kappa_{t, T}} \sim U[0.01, 0.50]$.}
  \label{fig_eps_050}
\end{subfigure}%
\begin{subfigure}{.5\textwidth}\centering
  \includegraphics[width=\textwidth]{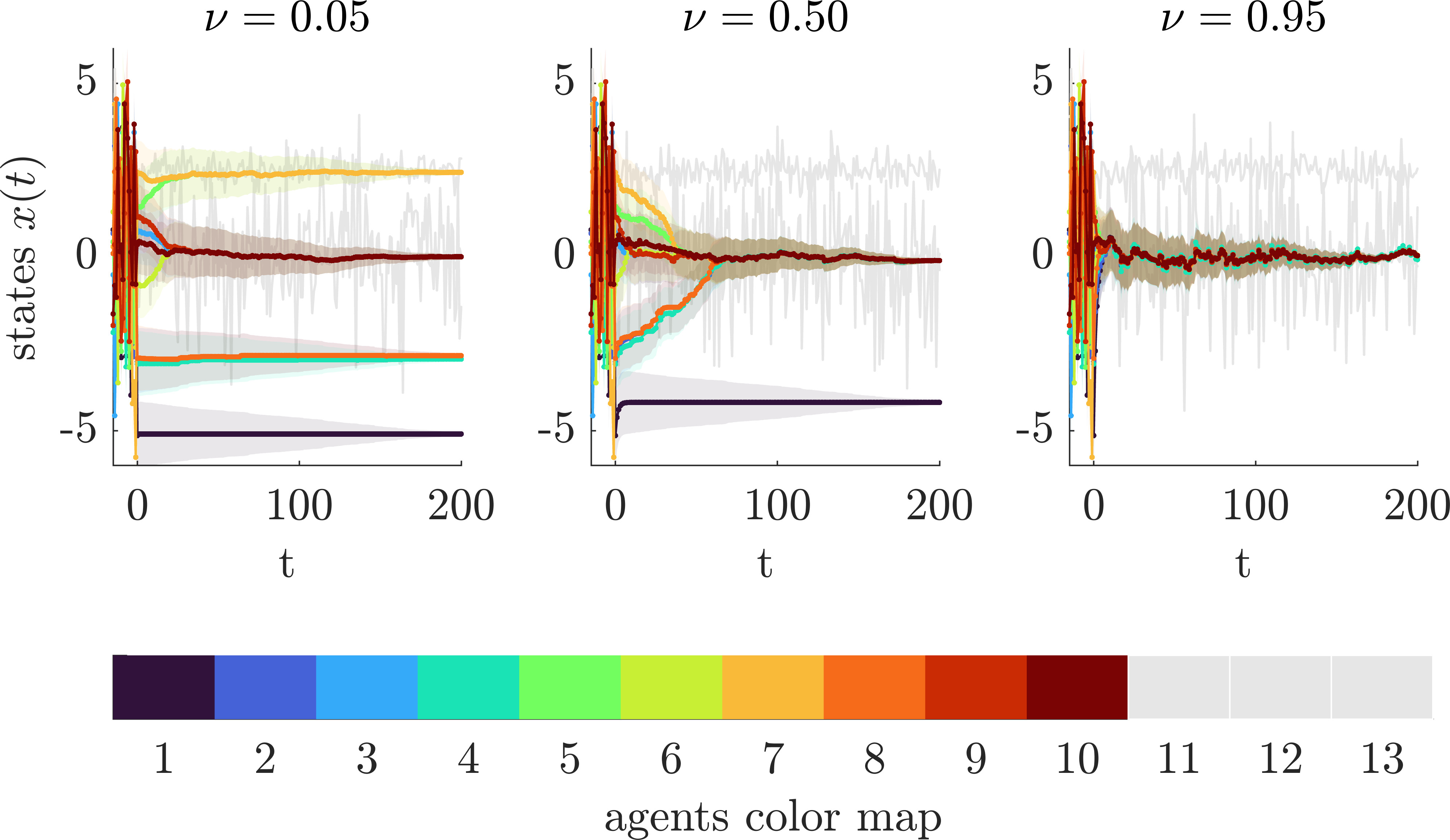}
  \caption{HDD protocol~\eqref{eqn_hdd_update} with $T = 15$, $\{\epsilon_{i,k}\}_{k \in \kappa_{t, T}} \sim U[0.01, 1.00]$.}
  \label{fig_eps_100}
\end{subfigure}\vspace{0.2cm}\\
\begin{subfigure}{.5\textwidth}\centering
  \includegraphics[width=\textwidth]{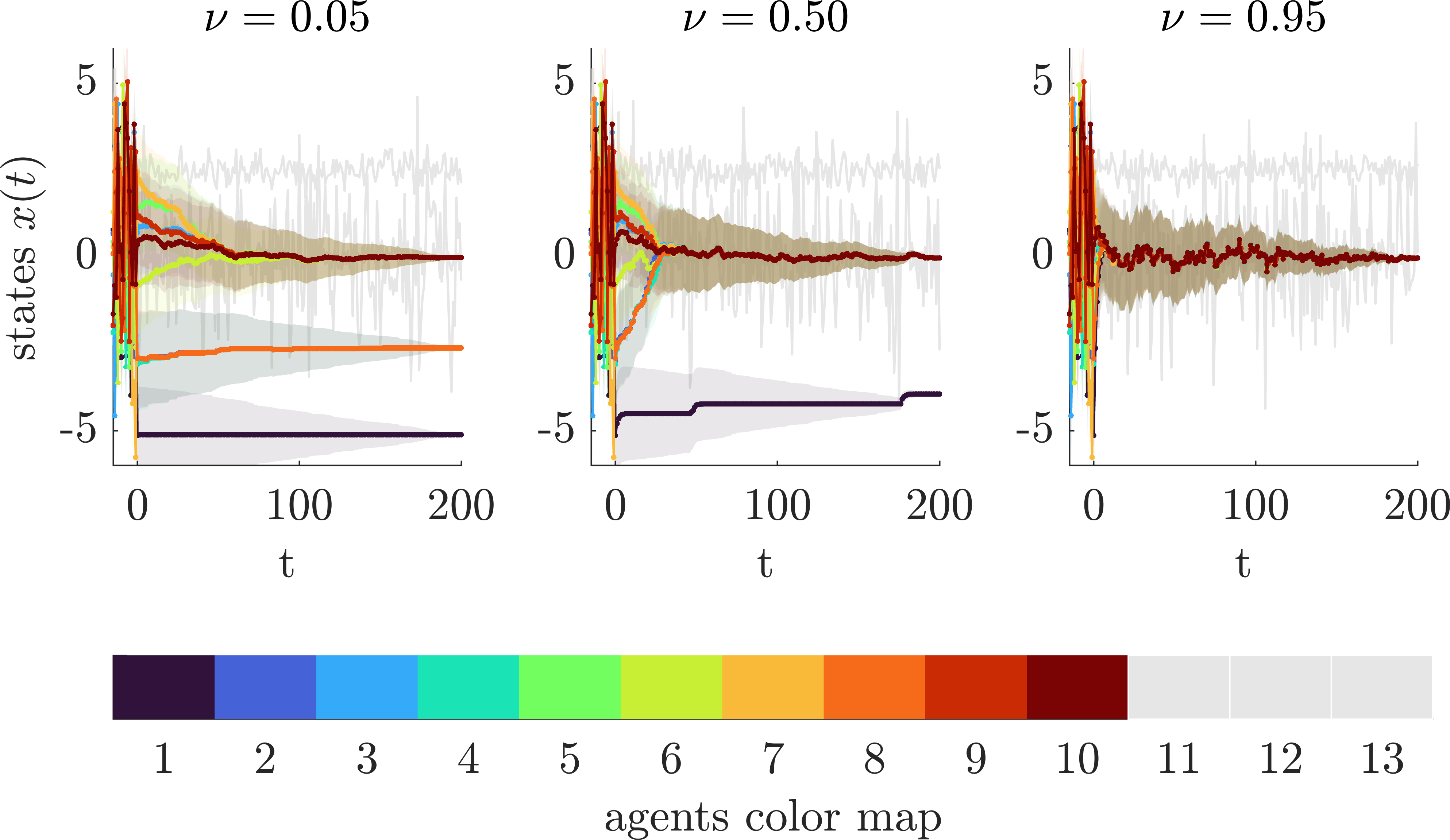}
  \caption{HDD protocol~\eqref{eqn_hdd_update} with $T = 15$, $\{\epsilon_{i,k}\}_{k \in \kappa_{t, T}} \sim U[0.01, 1.50]$.}
  \label{fig_eps_150}
\end{subfigure}%
\begin{subfigure}{.5\textwidth}\centering
  \includegraphics[width=\textwidth]{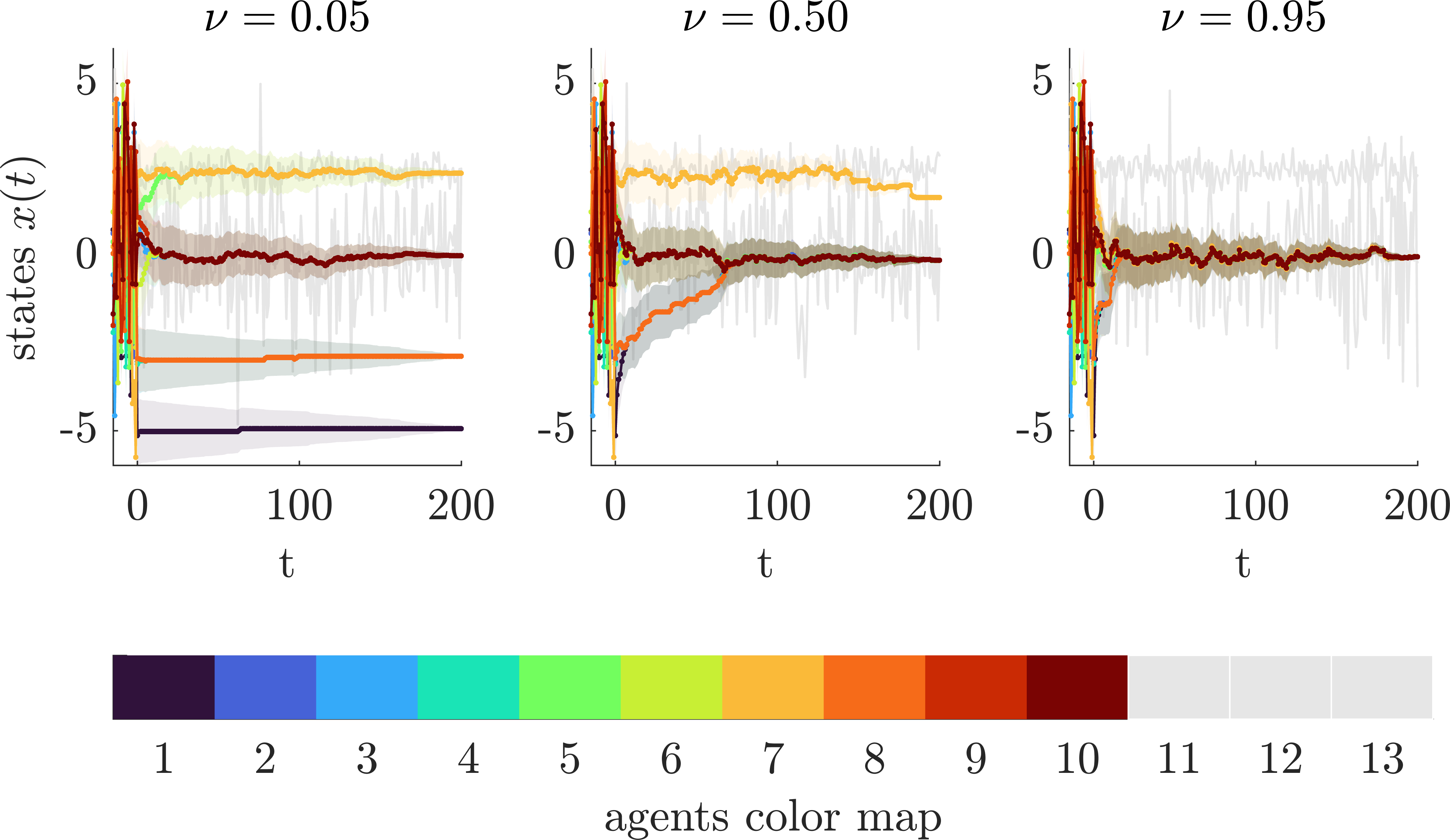}
  \caption{HDD protocol~\eqref{eqn_hdd_update} with $T = 5, \{\epsilon_{i,k}\}_{k \in \kappa_{t, T}} \sim U[0.01, 1.00]$.}
  \label{fig_T_5}
\end{subfigure}
\caption{Effect of parameters variation for the HDD protocol~\eqref{eqn_hdd_update}: states' trajectories $x(t)$. The color map indicates cooperative agents (i.e., $i \in \mathcal{V}_{c}$), while non-cooperative agents (i.e., $i \in \mathcal{V}_{nc}$) are shown in grey color. Each panel shows the evolution of agents' states $x_i(t)$, $i \in \mathcal{V}$, for $\nu=0.05,0.5,0.95$, with the confidence bounds $\{\epsilon_{i,k}\}_{k \in \kappa_{t, T}}\sim U[\underline{\epsilon}, \overline{\epsilon}]$ represented by shaded areas.
(a),(b),(c): Effect of varying the confidence bounds; here $T=15$, $\underline{\epsilon} = 0.01$ and $\overline{\epsilon} = \{0.5, 1.0, 1.5\}$. (d): Effect of varying the history length $T$; here $T=5$, $\underline{\epsilon} = 0.01$ and $\overline{\epsilon} = 1$.}
\label{fig_sim_results}
\end{figure*}

The results of our simulation are shown in Figure~\ref{fig_sim_results}. It is assumed that every non-cooperative agent follows a random state update rule. 
On all the sub-figures of Figure \ref{fig_sim_results}, the discount factor variations with $\nu \in \{0.05, 0.50, 0.95\}$ are shown. 
With low values of the discount factor, we observed clustering behaviour between agents, and with higher values of discount factor, the normal consensus convergence is observed. This is due to the fact that higher values of $\nu$ enabled the agents to remember the past interactions of their neighbors to a greater extent. 
The effect of varying the confidence bounds are shown in sub-figures \ref{fig_eps_050}, \ref{fig_eps_100} and \ref{fig_eps_150}, respectively for $\overline{\epsilon} = 0.5, 1.0, 1.5$. 
Higher values of confidence bounds encouraged the agents to cooperate with each other, while lower values of confidence bounds resulted in delayed cooperation and in clustering behaviour between agents.
Finally, when the history length was reduced to $T = 5$, the agents converged quickly thanks to small memory and big enough confidence bound. If the agents were to safeguard themselves against non-cooperative agents, their best bet would be to have small confidence bounds; however, such a strategy might lead to clustering behaviour. This explains that there is a definite trade-off that the agents have to observe if they plan on safely interacting with their neighbors. A detailed investigation of the effects of the parameters on the resulting HDD protocol weights and the final consensus value is available in the appendix of \cite{venki_angela_necsys_2022}. The code used to obtain the simulation results is made publicly available at \url{https://github.com/venkatramanrenganathan/HDDConsensus}.

\section{Conclusion \& Future Outlooks} \label{sec_conclusion}
We proposed a novel historical data-driven distributed consensus protocol for uncertain networks. Our proposed approach formulates the uncertainty about the trustworthiness of neighbors of an agent as a random vector termed as configuration, and learns the parameters defining its unknown but true distribution via history data-driven approach. Subsequently, the trustworthiness of all neighbors of an agent is inferred leading to the proposed HDD distributed consensus protocol. Our simulation results demonstrated the effectiveness of our proposed idea. As a future work, we seek to investigate the moment uncertainty along with the losses due to mistakenly associating wrong trust with neighbors given their historical data using distributionally robust optimization techniques. Other promising directions are investigating adaptive designs for confidence bounds and discount factors, and to design an history data-driven predictive consensus algorithm. 

\bibliography{ifacconf.bib}  

\newpage
\newpage
\onecolumn
\appendix
\section{Parameter Variations on HDD Protocol} \label{sec_appendix}
\begin{figure*}[h]
\begin{subfigure}{.5\textwidth}\centering
  \includegraphics[width=\textwidth]{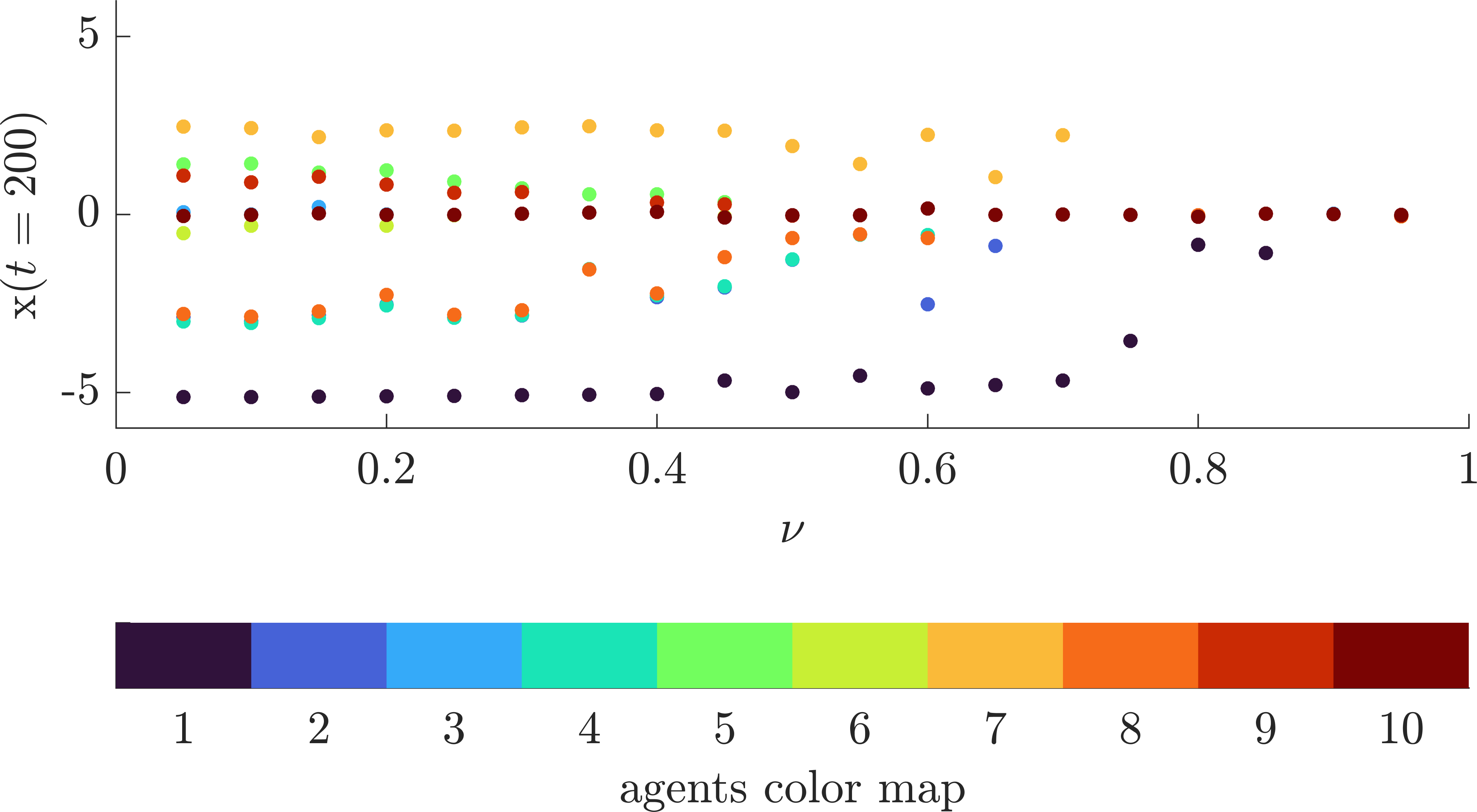}
  \caption{HDD protocol~\eqref{eqn_hdd_update} with $T = 15$, $\{\epsilon_{i,k}\}_{k \in \kappa_{t, T}} \sim U[0.01, 0.50]$.}
  \label{fig_endeps_050}
\end{subfigure}%
\begin{subfigure}{.5\textwidth}\centering
  \includegraphics[width=\textwidth]{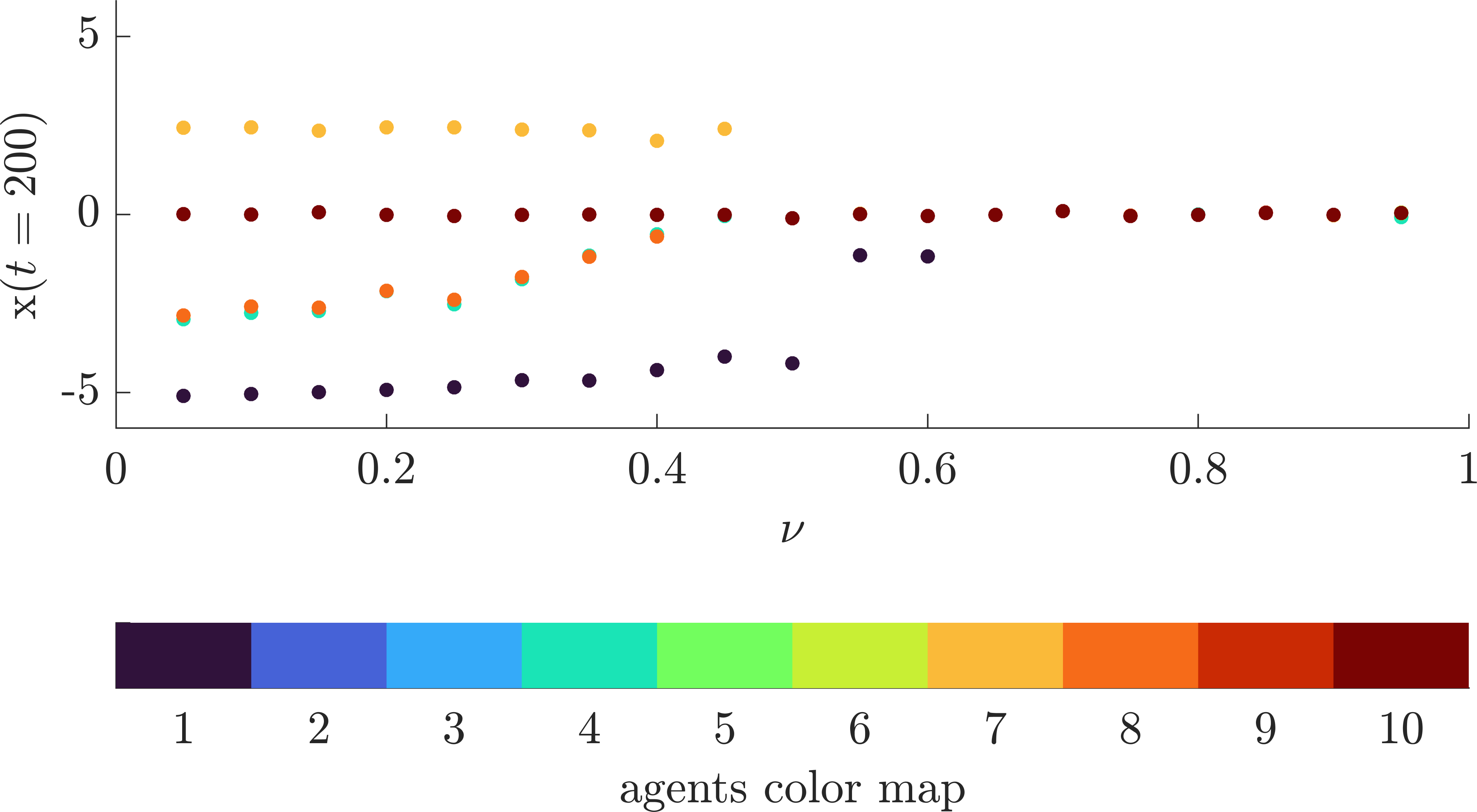}
  \caption{HDD protocol~\eqref{eqn_hdd_update} with $T = 15$, $\{\epsilon_{i,k}\}_{k \in \kappa_{t, T}} \sim U[0.01, 1.00]$.}
  \label{fig_endeps_100}
\end{subfigure}
\begin{subfigure}{.5\textwidth}\centering
  \includegraphics[width=\textwidth]{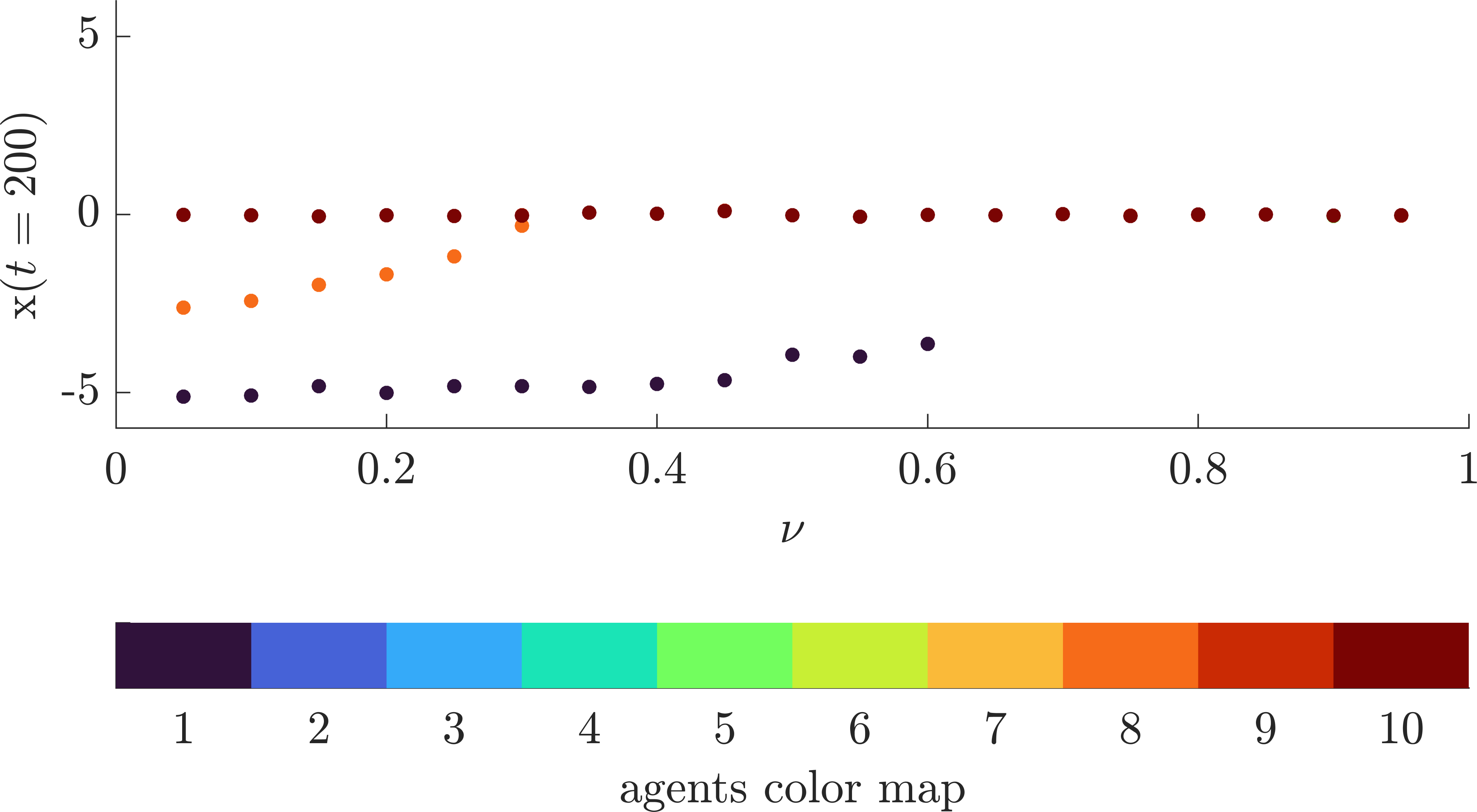}
  \caption{HDD protocol~\eqref{eqn_hdd_update} with $T = 15$, $\{\epsilon_{i,k}\}_{k \in \kappa_{t, T}} \sim U[0.01, 1.50]$.}
  \label{fig_endeps_150}
\end{subfigure}%
\begin{subfigure}{.5\textwidth}
  \centering
  \includegraphics[width=\textwidth]{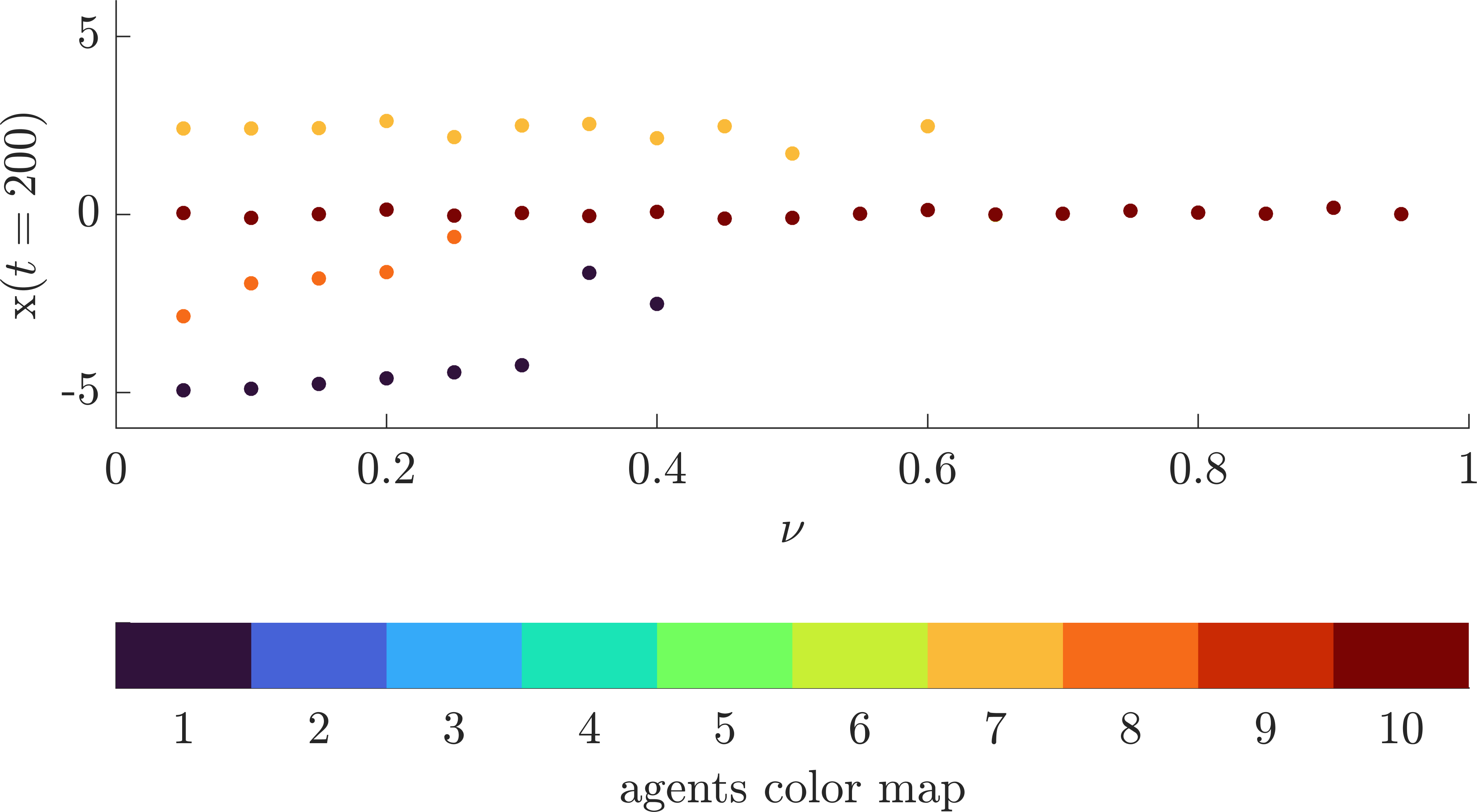}
  \caption{HDD protocol~\eqref{eqn_hdd_update} with $T = 5$, $\{\epsilon_{i,k}\}_{k \in \kappa_{t, T}} \sim U[0.01, 1.00]$.}
  \label{fig_endT_5}
\end{subfigure}
\caption{
Effect of parameters variation for the HDD protocol~\eqref{eqn_hdd_update}: agreement vs. clustering.
The color map indicates cooperative agents (i.e., $i \in \mathcal{V}_{c}$). Each panel shows the states of cooperative agents $x_i(t)$, $i \in \mathcal{V}_c$, at time $t=200$ for increasing values of $\nu\in \{0.05,0.1,\dots,0.95\}$, with confidence bounds $\{\epsilon_{i,k}\}_{k \in \kappa_{t, T}}\sim U[\underline{\epsilon}, \overline{\epsilon}]$. Clustering and consensus behaviour is consistently observed on all four subplots with respect to lower discount factor and higher discount factor settings respectively.}
\label{fig_endx_results}
\end{figure*}

\begin{figure*}
\begin{subfigure}{.5\textwidth}
  \centering
  \includegraphics[width=\textwidth]{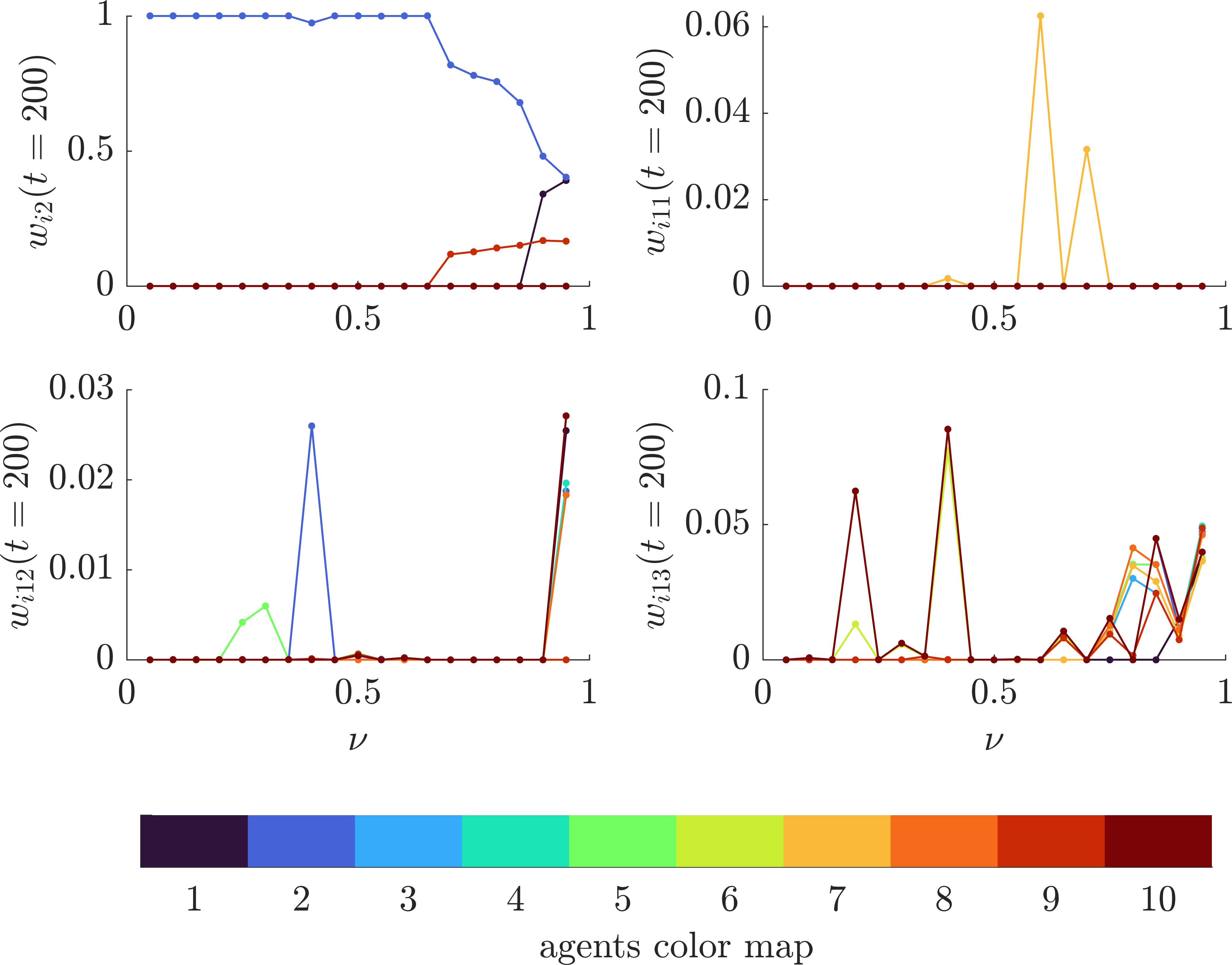}
  \caption{HDD protocol~\eqref{eqn_hdd_update} with $T = 15$, $\{\epsilon_{i,k}\}_{k \in \kappa_{t, T}} \sim U[0.01, 0.50]$.}
  \label{fig_wendeps_050}
\end{subfigure}%
\begin{subfigure}{.5\textwidth}
  \centering
  \includegraphics[width=\textwidth]{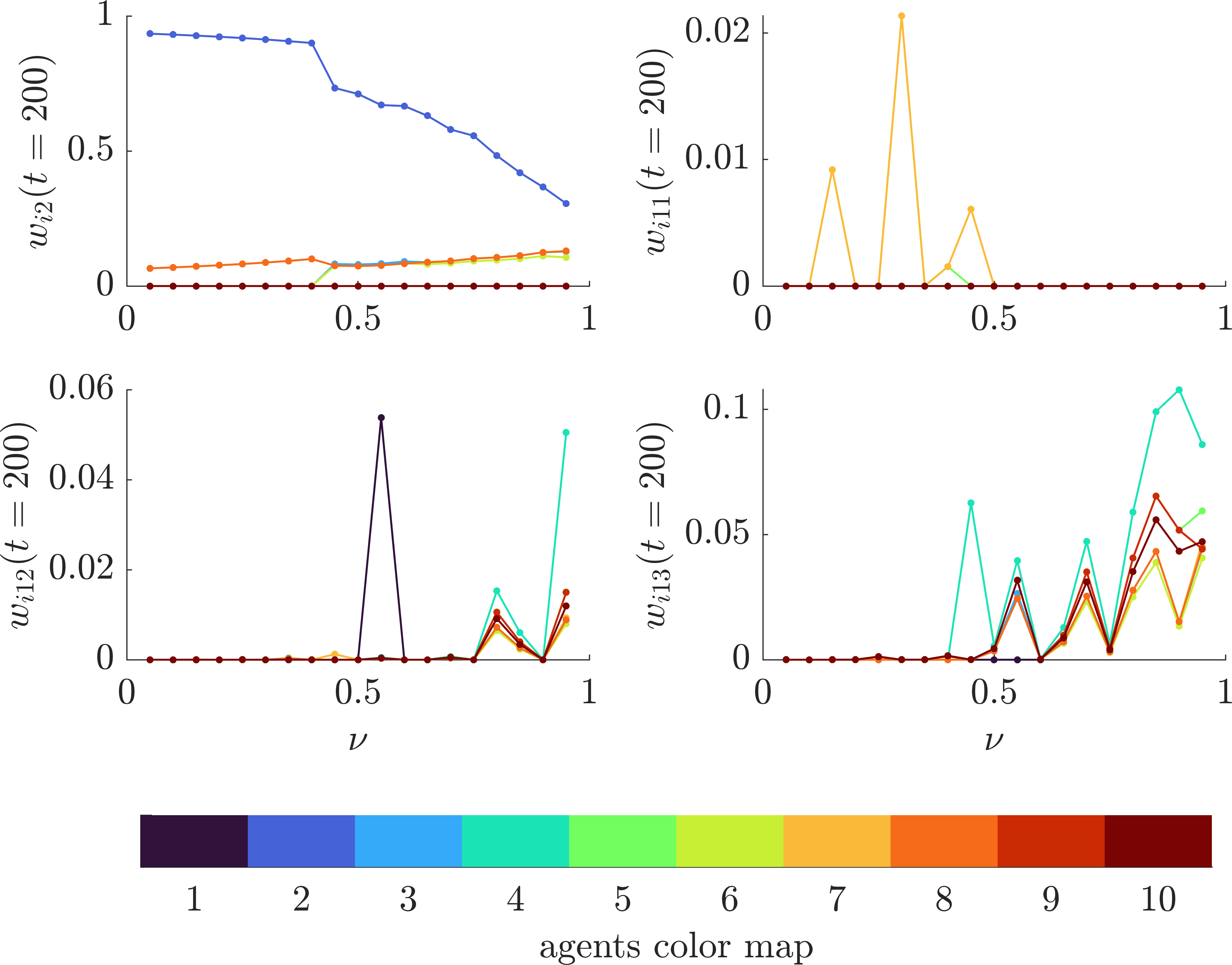}
  \caption{HDD protocol~\eqref{eqn_hdd_update} with $T = 15$, $\{\epsilon_{i,k}\}_{k \in \kappa_{t, T}} \sim U[0.01, 1.00]$.}
  \label{fig_wendeps_100}
\end{subfigure}\vspace{0.2cm}\\
\begin{subfigure}{.5\textwidth}
  \centering
  \includegraphics[width=\textwidth]{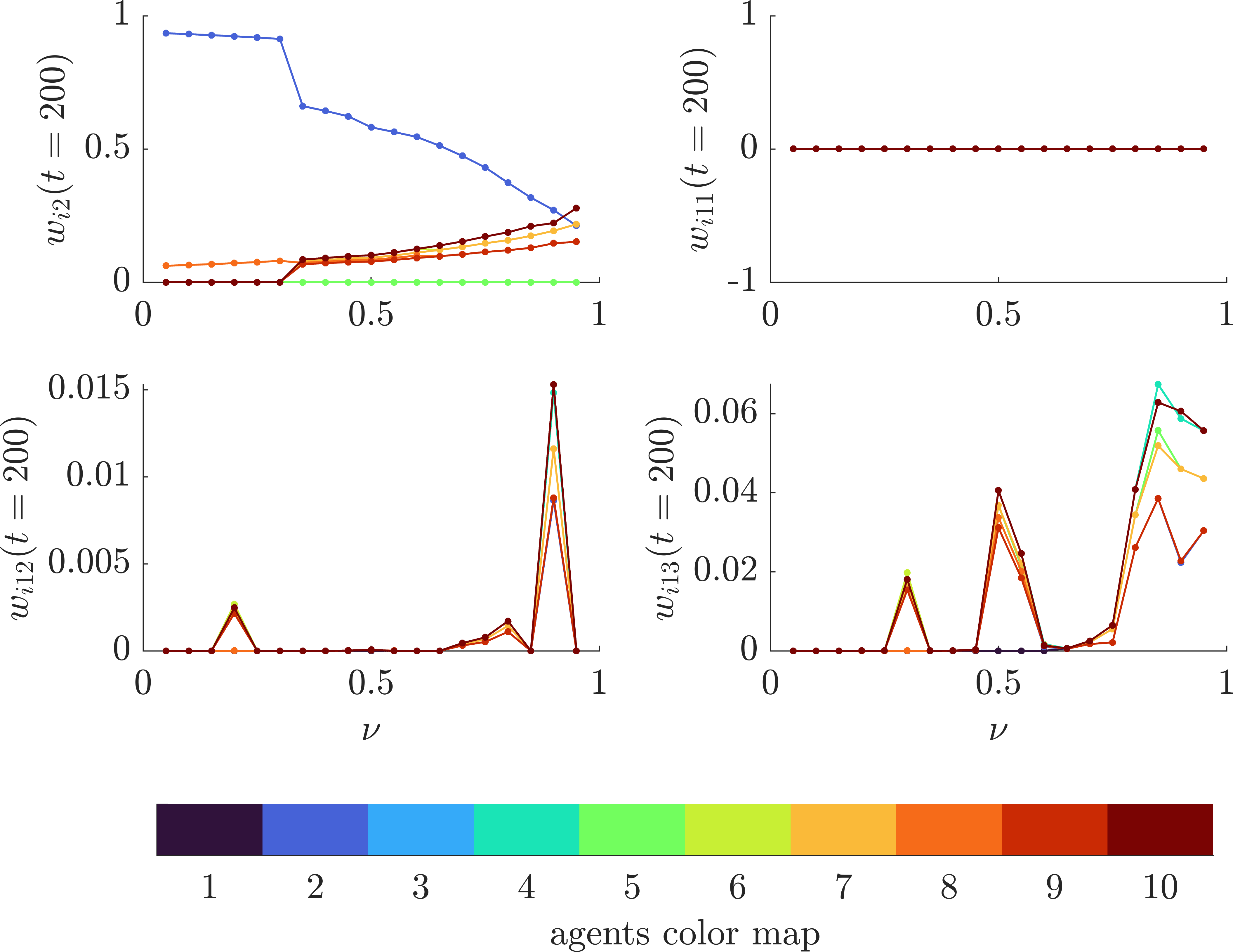}
  \caption{HDD protocol~\eqref{eqn_hdd_update} with $T = 15$, $\{\epsilon_{i,k}\}_{k \in \kappa_{t, T}} \sim U[0.01, 1.50]$.}
  \label{fig_wendeps_150}
\end{subfigure}%
\begin{subfigure}{.5\textwidth}
  \centering
  \includegraphics[width=\textwidth]{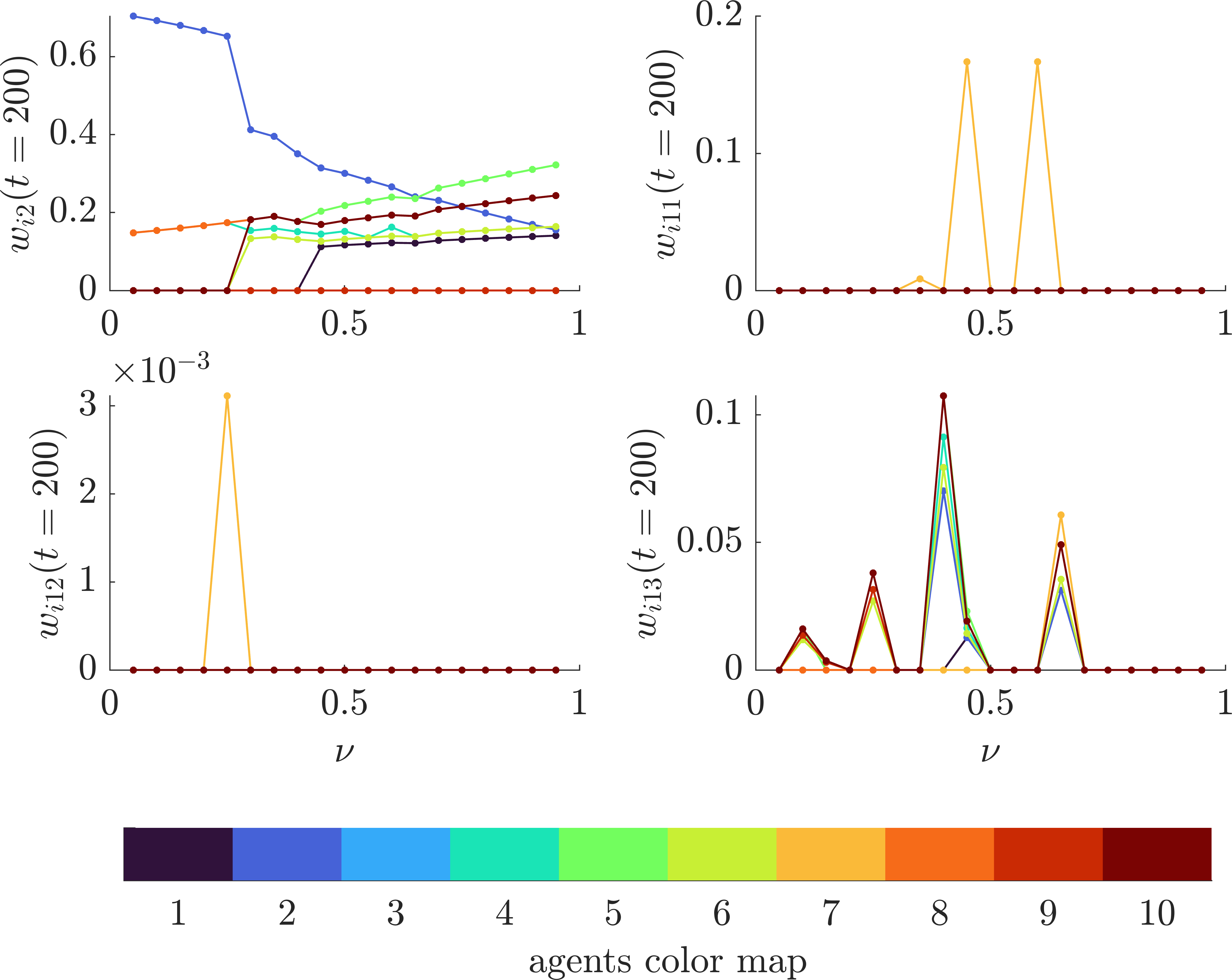}
  \caption{HDD protocol~\eqref{eqn_hdd_update} with $T = 5$, $\{\epsilon_{i,k}\}_{k \in \kappa_{t, T}} \sim U[0.01, 1.00]$.}
  \label{fig_wendT_5}
\end{subfigure}
\caption{Effect of parameters variation for the HDD protocol~\eqref{eqn_hdd_update}: matrix $W(t)$. The color map indicates cooperative agents (i.e., $i \in \mathcal{V}_{c}$). Each panel shows the elements $w_{ij}(t)$ for $i \in \mathcal{V}_c$ (i.e, cooperative agents) and $j=2,11,12,13$ (top-left top-right, bottom-left, bottom-right, respectively) at time $t=200$, for increasing values of $\nu\in \{0.05,0.1,\dots,0.95\}$, with confidence bounds $\{\epsilon_{i,k}\}_{k \in \kappa_{t, T}}\sim U[\underline{\epsilon}, \overline{\epsilon}]$.}
\label{fig_endw_results}
\end{figure*}

\begin{figure}
    \centering
    \includegraphics[scale=0.125]{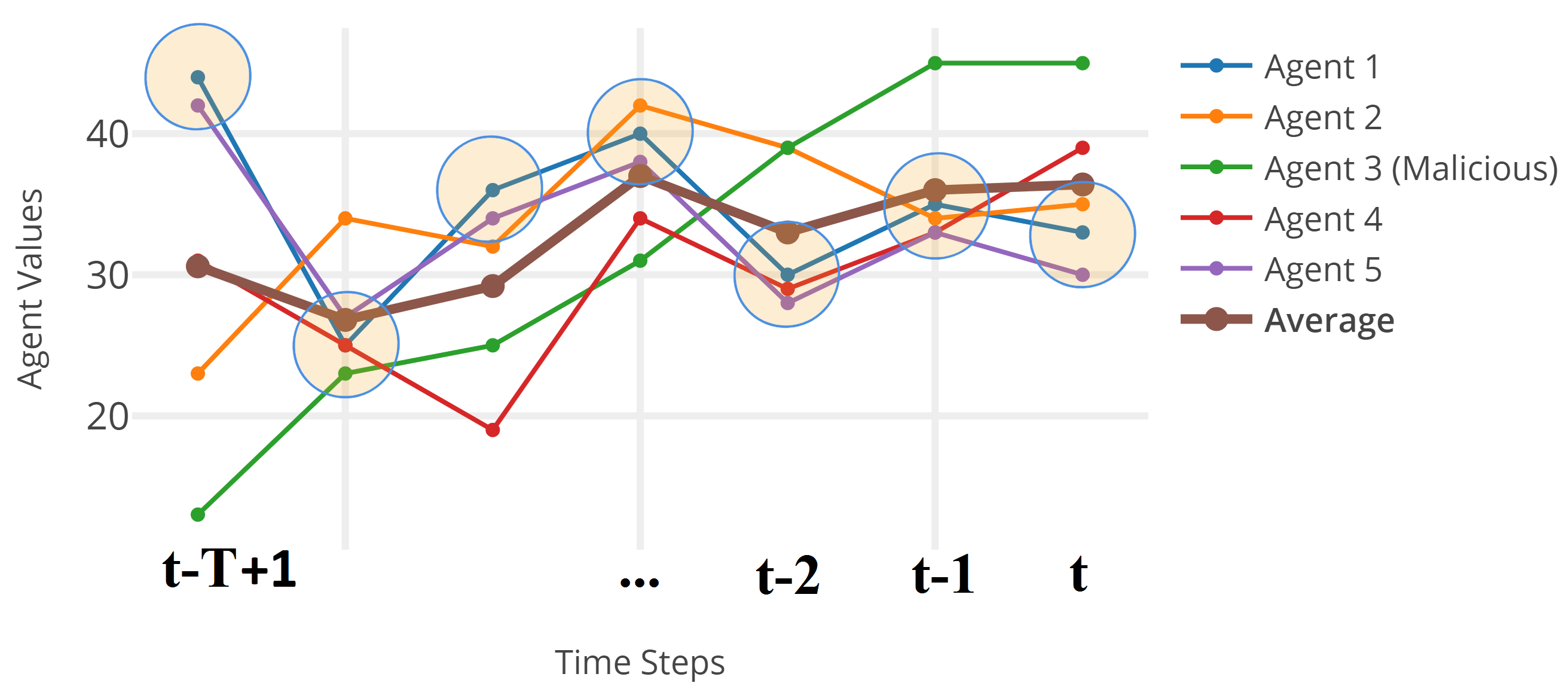}
    \caption{An $\epsilon$-neighborhood based set membership for agent $\mathbf{1}$, namely $\mathcal{B}_{x_{1}(k)}(\epsilon_{i, k})$ with $\epsilon_{i, k} = \epsilon > 0, \forall k \in \kappa_{t, T}$ corresponding to the past $T$ time steps historical data is illustrated here. In our work, we propose to have the $\epsilon$ balls to be decreasing when time moves forward along with other parameters.}
    \label{fig_set_membership}
\end{figure}

A basic illustration of our approach with constant $\epsilon_{i,k}, k \in \kappa_{t,T}$ is shown in Fig.~\ref{fig_set_membership}. Now, consider the same example presented in section \ref{sec_numerical_example} as shown in Fig.~\ref{fig_sim_results}. The resulting trust-based consensus or (clustering) value at time $t = 200$ is plotted in Fig.~\ref{fig_endx_results}. When the forgetting factor was small in addition to the confidence bounds $\epsilon_{t}$ being small, we observed a ``clustering'' behavior: this is due to the fact that the HDD protocol values only the most recent past. For instance, for $\nu < 0.1$, $\nu^{t-k}\approx 0$ for $k=t-T+1,\dots,t-2$, that is, the agents ``forget quickly''. Instead, when the forgetting factor is close to $1$, the states tend to converge to consensus, due to the ability of the agents to remember past events (such as remembering agent $j$ in its $\epsilon_k$-neighborhood for a certain time $k\in\{t-T+1,\dots,t\}$).
Fig.~\ref{fig_endw_results} shows the elements of the $j^\text{th}$ ($j=2,11,12,13$) column of $W(t)$ when $t=200$, for increasing values of $\nu$. Remember that each element $w_{ij}$ ($i\in \mathcal{V}_{c}, j=2,11,12,13$) represents the trust that each agent $i$ has on its neighbor $j$; then, the intuition is that if there exists a value of $\nu$ such that $w_{ij}(200) = 0$ for all $i\in \mathcal{V}_{c}$ and $j \in \mathcal{V}_{nc}$, it means that the cooperative agents correctly decide to not trust the non-cooperative agents. For instance, Fig.~\ref{fig_endw_results} shows that the non-cooperative agents $11,12$ are ``detected'' for most values of $\nu$ (except for agent $7$ in yellow which almost always believes the non-cooperative neighbors $(j \in \mathcal{V}_{nc})$ no matter what values of confidence bounds and discount factor are used). One could think that $\nu\approx 1$ corresponds to an optimal choice; this is however not the case if a non-cooperative agent adopts a smart behavior (see agent $13$ and bottom-right panel of all sub-figures in Fig.~\ref{fig_endw_results}) and obtains the trust of all the other agents. Finally, it is interesting to notice that in order to achieve cooperation, an agent needs to trust its neighbors' states and lower the certainty in its own state (see the element $w_{22}(200)$ depicted in blue in the top-left panel of all four sub-figures in Fig.~\ref{fig_endw_results}). Specifically, a cooperative agent (like agent $2$) under smaller confidence bounds is stubborn initially with smaller discount factor and then relaxes its certainty upon itself with higher discount factor resulting in consensus. On the other hand, when the confidence bounds increased, the rate at which it relaxes its certainty increases and, as a result, a faster cooperation is observed. Under the effect of different history lengths depicted by the top right and bottom right sub-figures of Fig.~\ref{fig_endw_results}, we see that under shorter history length, the cooperative agents relax a lot faster their certainties leading to faster cooperation as their memory is smaller. Note that if the cooperative agents were to relax their certainty and cooperate faster, it might come at the expense of \emph{potentially} starting to believe non-cooperative neighbors. Future work will seek to design an adaptive sequence of confidence bounds and discount factor at each time step to rectify this phenomenon and encourage agents to include neighbors with more distant opinions.

\end{document}